\documentclass[aip, jmp, amsmath,amssymb,preprint, reprint, author-year, author-numerical]{revtex4-1}

\usepackage{graphicx}% Include figure files
\usepackage{dcolumn}% Align table columns on decimal point
\usepackage{bm}% bold math
\usepackage{amsmath}
\usepackage{amsfonts}
\usepackage{color}
\usepackage{subfigure}
\usepackage{listings}
\usepackage{caption}
\usepackage{mathrsfs}

\begin{document}

\preprint{AIP/123-QED}

\title[Correlation of Dirac potentials and atomic inversion in cavity quantum electrodynamics]{Correlation of Dirac potentials and atomic inversion in cavity quantum electrodynamics}% Force line breaks with \\
%\thanks{Footnote to title of article.}

\author{Agung Trisetyarso}
\email{trisetyarso07@a8.keio.jp}
\affiliation{School of Fundamental Science and Technology, Keio University\\
3-14-1 Hiyoshi, Kohoku-ku, Yokohama-shi, Kanagawa-ken 223-8522, Japan.}%Lines break automatically or can be forced with \\

\date{\today}% It is always \today, today,
             %  but any date may be explicitly specified

\begin{abstract}
Controlling the time evolution of the population of two states in cavity quantum electrodynamics (QED) is necessary by tunning the \textit{modified} Rabi frequency in which the $\textit{extra}$ classical effect of electromagnetic field is taken into account.The theoretical explanation underlying the perturbation of potential on spatial regime of bloch sphere is by the use of (Bagrov, Baldiotti, Gitman, and Shamshutdinova) BBGS-Darboux transformations \cite{bagrov14darboux} on the electromagnetic field potential in one-dimensional stationary Dirac model in which the Pauli matrices are the central parameters for controlling the collapse and revival of  the Rabi oscillations. It is shown that by choosing $\sigma_{1}$ in the transformation generates the parabolic potential causing the total collapse of oscillations; while  $\{\sigma_{2}, \sigma_{3}\}$ yield the  harmonic oscillator potentials ensuring the coherence of qubits.

\end{abstract}

% PACS, the Physics and Astronomy
                             % Classification Scheme.
%Use showkeys class option if keyword
                              %display desired
\maketitle

\section[INTRODUCTION]{\label{sec:level1}INTRODUCTION}

Recent research on quantum theory scrutinizes the reconciliation between quantum and classical physics, and it attempts to take the advantages from the classical for constructing the new theories \cite{PhysRevD.47.3345}.  For instance, the quantum information between sender and receiver can be teleported by involving classical information transmission generated by the joint measurement of EPR pair \cite{PhysRevLett.70.1895}. In the modern research of experimental quantum information, the notions of Coulomb and photon blockades represent a recent resurgence of interest in the influence of classical in quantum physics \cite{PhysRevLett.79.1467,birnbaum2005photon}. Here, we show that concurrent condition of the electromagnetic field behaving classically and quantumly in cavity QED provides the conditions for steering the dynamics of collaps and revival sequences of Rabi oscillations.

Rabi oscillations are the important features in the field of experimental quantum information, since the oscillations represent the coherence and decoherence of the qubits in various type of physical schemes of quantum computer\cite{mooij1999josephson,PhysRevLett.76.1800,stievater2001rabi,boehme2003electrical}.  

The famous Rabi oscillations are derived from the quantity associated as the atomic inversion, $W(t)$, defined as the variation in the excited- and ground-state populations \cite{knight2005introductory}
\begin{equation}
\label{eq0a}
W(t)=P_{e}(t)-P_{g}(t).
\end{equation}
When the electromagnetic field is classical, the oscillations in atomic inversion have constant amplitude. If the field is quantum, the oscillations consist of a sequence of collapses and revivals. The object of the control of the time evolution of the population of two states in cavity is to formulate the Hamiltonian $H$, which realizes the population quantity of excited- and ground-state \cite{H.Mabuchi11152002}.

In the recent years, there have been several efforts to control the collapses and revivals \cite{PhysRevLett.92.223004,PhysRevA.33.3610,flagg2009resonantly,rudin2001anharmonic}, in order to damp out \cite{PhysRevLett.91.127401,PhysRevB.72.172509,PhysRevA.29.2627} or to ensure the stability \cite{shao2004dynamics} of the oscillations by considering the trade-off of physical constraints in the related quantum computer scheme. Nevertheless, the general theory of how to control the atomic inversion is poorly comprehended. 

On the theoretical side, Darboux transformations contributions into the wide range of physics fields have been widely known over centuries \cite{matveev1991darboux}. Different with the other transforms such as the Fourier and Laplace transforms, the action of Darboux transform does not change the \textit{domain} of the potential and state of the system, but it creates the new potential and state of the system. The output of Darboux transform is a set of new potential and state of the physical system. For instance, in nonlinear Schr\"{o}dinger equation, Darboux transform generates the new solutions of the equation \cite{matveev1991darboux}. The Darboux transform is very useful to describe the evolution of a physical system due to the change of its Hamiltonian potential as found in wide range of nonlinear physics. However, its contribution to the field of quantum computation is still poorly understood.

Recently, there are several efforts in order to open the mystery of Darboux transformations into quantum computation. Samsonov $et$ $al.$, followed by Hussin $et$ $al.$ \cite{hussin2006generalized}, propose a novel technique to apply Darboux transformations into Jaynes-Cumming Hamiltonian \cite{0305-4470-37-43-007}. Bagrov $et$ $al.$ coined a Darboux transformation which does not violate the two-level system equation structure \cite{bagrov14darboux}. These processes can be done by using the method of intertwining on the exactly solvable Dirac potentials which are equivalent to the Darboux transformations between Schr\"odinger potentials \cite{PhysRevA.43.4602}. 

The application of Darboux transformations to one-dimensional Dirac equation is performed by intertwining the Hamiltonian in Dirac equation
\begin{equation}
\label{eqdh}
\hat{L}h_{0}=h_{1}\hat{L},
\end{equation}
where $\hat{L}$ is intertwining operator and $h_{0}$ ($h_{1}$)  is the old (new) Dirac Hamiltonian. In the sense of perturbation theory, $h_{0}$ ($h_{1}$) is unperturbed (perturbed) Dirac Hamiltonian, where $h_{1}=h_{0}+V(\alpha,~\beta)$, in which $\{\alpha,~\beta\}$ contributing into the perturbation terms. Recently, it has been shown that numerous problems in quantum mechanics can be solved under this technique \cite{Nieto2003151}.

To date, the closest theoretical result to the experimental was only successfully performed by Samsonov $et$ $al.$ by showing the contribution of pseudosymmetry potential in controlling the qubits dynamics of superconducting circuits based on Josephson tunnel junction\cite{1751-8121-41-24-244023}. In this paper, we would like to show that it is possible to control Rabi oscillations in cavity QED by considering the presence of extra classical electromagnetic field. It emerges the perturbation effect on the system and can be represented by one fold Darboux transformations on one-dimensional Dirac equation. The results suggest that there is a correlation of Dirac potentials and the atomic inversion.

The paper is organized as follows : in Subsection (\ref{level1r}), we introduce the method proposed by Bagrov $et$ $al.$, then we generalized the methods : Pauli matrix in operator $B$ is not constrained in $\sigma_{3}$, but the use of $\{\sigma_{1},\sigma_{2}\}$ is also considered. Moreover, the potential is not only restricted in (pseudo)-scalar form, but also we allow the potential in the form of vector where the Pauli matrices act as the potential basis and the function as the potential coefficient. Subsection (\ref{level1d}) contains the introduction of the mathematical model of atom-field interaction in cavity QED. Here, the Rabi frequency and oscillations of fully-quantum mechanical model are modified due to the presence of the $\textit{extra}$ classical effect of electromagnetic field. Section (\ref{level2aw}) provides the elucidation of the implementation of BBGS-Darboux transformations to semiclassical Rabi model. The outcome is to obtain the parameters in BBGS-Darboux transformations : $\alpha(t)$, $\beta(t)$, and $|V(t)|$. The section covers two types of procedures: \textit{first}, reformulation of the equation from Schr\"{o}dinger equation into Dirac equation. Here, the representation of data is changed, but we do not control anything. \textit{Second}, the action of Darboux transformation on the Dirac equation. In this step, we perturb the system to control the potential and state of the system. In Section (\ref{level6az}), it is shown that the choice
 of Pauli matrices in BBGS-Darboux transformations relates to the type of  one-dimensional stationary Dirac potential of electromagnetic field : $\{\sigma_{1}\}$ generates the parabolic potential (so-called $\textit{Yukawa-Rabi~oscillations}$) and $\{\sigma_{2},\sigma_{3}\}$ yields the  harmonic oscillator potential. 

\subsection[BBGS-Darboux transformations]{\label{level1r}BBGS-Darboux transformations}

In this Subsection, we introduce (Bagrov, Baldiotti, Gitman, and Shamshutdinova) BBGS-Darboux transformations \cite{bagrov14darboux} which has the notation ${\mathcal D(\sigma_{i})}$. The symbol $\sigma_{i}$ is the Pauli matrix in element of intertwining operator $\hat{L}=A\frac{d}{dt}+B$, $A=\sigma_{0}$=$( \begin{smallmatrix}1& 0 \\0& 1\end{smallmatrix} )$, and $B=\alpha_{i}(t)+(f(t)-\beta_{i}(t))\sigma_{i}$.

The action of  ${\mathcal D(\sigma_{i})}$ on  a set of potential and state of a Hamiltonian is defined by ${\mathcal D (\sigma_{i})}[N]$ $\{V,\Psi\}$ = $\{V[N],\Psi[N]\}$ transforming the old Hamiltonian $\hat{h}(V)\Psi =\varepsilon_{0}\Psi$ $\rightarrow$ $\hat{h}(V[1])\Psi[1] =\varepsilon_{1}\Psi[1]$ $\rightarrow$ ... $\rightarrow$  $\hat{h}(V[N])\Psi[N] =\varepsilon_{N}\Psi[N] $. The potential is a vector in three dimensional Euclidean Bloch sphere using Pauli matrices as the basis and it can be extended into \textit{n}-dimensions \cite{Trisetyarso:arXiv1003.4590}. The transformations affect the potential from $V(t)=\sum_{j}\sigma(j)^{j}f_{j}(t)$ $\rightarrow$ $V[1](t)=\sum_{j}\sigma(j)^{j}f'_{j}(t)$, where $f'_{j}(t)=f_{j}(t)+\triangle f_{j}(t)$ and $\sigma(j)\in \{\sigma_{1}, \sigma_{2}, \sigma_{3}\}$. Based on this formulation, the magnitude of potential is transformed $|V(t)|=\sum_{j}|f_{j}(t)|$ $\rightarrow$ $|V_{i}[N](t)|=\sum_{j}|f_{j}'(t)|$. The effect of the transformations to the state is $\Psi$ $\rightarrow$ $\hat{L}\Psi=\Psi[N]$. In order to obtain the coefficients $\alpha_{i}(t)$ and $\beta_{i}(t)$, intertwining operation, $\hat{L}\hat{h}_{\text{old}}=\hat{h}_{\text{new}}\hat{L}$, between the old and the new Hamiltonian is needed. For $N$-fold BBGS-Darboux transformations, $\hat{L}\hat{h}_{\text{0}}=\hat{h}_{\text{1}}\hat{L}$ $\rightarrow$ $\hat{L}\hat{h}_{\text{1}}=\hat{h}_{\text{2}}\hat{L}$  $\rightarrow$ ...  $\rightarrow$ $\hat{L}\hat{h}_{\text{N-1}}=\hat{h}_{\text{N}}\hat{L}$.

After the above definitions, we can now express the following theorem.

\textit{${\textbf {Theorem 1}}$. Let $\{V,\Psi\}$ represents a physical system and ${\mathcal D (\sigma_{i})}[N]$ 
is a BBGS-Darboux transformation operator. The open-loop control mechanism can be constructed, where the BBGS-Darboux transformation operator can be assumed as a controller, $\{V,\Psi\}$ as the initial system, $\{V[N],\Psi[N]\}$ as the final system, and the eigenvalues $\varepsilon_{N}(\sigma_{i})$ are the output variables.} 

$\textit{Proof}$.  Let  $\{V,\Psi\}$ is the initial condition of a physical system, in which the initial eigenvalues $\varepsilon_{0}$ are belong to the system. 

The action of BBGS-Darboux transformation on this system is ${\mathcal D (\sigma_{i})}[N]$ 
 $\{V,\Psi\}$ =  $\{V[N],\Psi[N]\}$, in which the eigenvalues $\varepsilon_{N}(\sigma_{i})$ are belong to the new eigenstates $\Psi[N]$.

The complete proofs of the above theorem are given in Section \ref{level2aw} and Section \ref{level6az}. In this paper, we only consider $N=1$ or one fold Darboux transformations.

\subsection[Jaynes-Cummings model]{\label{level1d}Cavity QED model}

Following the Ref.   \cite{babelon2007short,knight2005introductory}, the system of  a two level atom and photons in a single cavity mode, if the electromagnetic field is classical, is represented by $H=\hbar \omega_{0} \frac{\sigma_{z}}{2}+\hbar \Omega[\sigma^{+}b(t)+\sigma^{-}b(t)^{\dagger}]$ where $b(t)=b_{0}e^{-i\omega t}$. If the electromagnetic field is quantum, the Hamiltonian is $H=\hbar \omega_{0} \frac{\sigma_{z}}{2}+\hbar \omega {\mathcal B}^{\dagger}{\mathcal B}+\hbar \Omega[\sigma^{+}{\mathcal B}+\sigma^{-}{\mathcal B}^{\dagger}]$, where ${\mathcal B}$ (${\mathcal B^{\dagger}}$) is annihilation (creation) operator and $|\gamma|^{2}$ is the average photon number.

It is shown that the atomic inversion of Jaynes-Cummings Hamiltonian is represented by $W(t)$=$\frac{\langle\gamma,\uparrow|\sigma_{z}(t)|\gamma,\uparrow\rangle}{\langle\gamma,\downarrow|\sigma_{z}(0)|\gamma,\downarrow\rangle}$ = $e^{-|\gamma|^{2}}\sum_{n=0}^{\infty}\frac{|\gamma|^{2n}}{n!}(1-2\Omega^{2}(n+1)) \frac{sin^{2}(\Omega_{n}t)}{\Omega_{n}^{2}}$, where $\Omega_{n} = \sqrt{\Omega(\kappa^{2}+n+1)}$, $\kappa=\frac{\triangle}{2\Omega}$, and $\triangle = \omega_{0}-\omega$ is the resonance condition. In this paper, we use $\bar{n}$=30, $\triangle=2\sqrt{2}$, and $\Omega$=1.

For our purpose, we $\it{modify}$ the Rabi frequency by involving additional term representing the classical effect of electromagnetic field. The modified Rabi frequency is \begin{equation}\label{mrf}\Omega_{n}^{m}= \sqrt{\Omega(\kappa^{2}+b^{\dagger}b+n+1)}.\end{equation}  

The superscript `$\it{m}$' denotes the modification. The equation (\ref{mrf}) represents the condition that the electromagnetic field $\textit{concurently}$ behaves classically and quantumly : the term $\sqrt{b^{\dagger}b}$ corresponds to classical field and $\sqrt{n+1}$ is the consequence of quantum effect of the field. This term can be easily derived from a Hamiltonian driven by a classical light source as mentioned in Ref. \cite{walls2006quantum}. Another way, it also might be obtained by transforming Jaynes-Cummings Hamiltonian as explained in Ref. \cite{0305-4470-37-43-007}. Also, the \textit{modified} atomic inversion is 
\begin{eqnarray}
\label{modatin}
W(t)={\frac{\langle\gamma,\uparrow|\sigma_{z}(t)|\gamma,\uparrow\rangle}{\langle\gamma,\downarrow|\sigma_{z}(0)|\gamma,\downarrow\rangle}}^{m}= ~~~~~~~~~~~~~~~\nonumber
\\e^{-|\gamma|^{2}}\sum_{n=0}^{\infty}\frac{|\gamma|^{2n}}{n!}\bigl(1-2(\Omega)^{2}\sqrt{b^{\dagger}b+n+1}\frac{sin^{2}(\Omega_{n}^{m}t)}{(\Omega_{n}^{m})^{2}}\bigr).
\end{eqnarray}

Darboux transformation contributes to the $b^{\dagger}b$ term. For $\it{N}$-fold Darboux transformations, it changes $b^{\dagger}b$ $\rightarrow$ $b_{i}[1]^{\dagger}b_{i}[1]$ $\rightarrow$ ... $\rightarrow$ $b_{i}^{\dagger}[N]b_{i}[N]$, where subscript $i$ corresponds to the choice of Pauli matrix in the intertwining operator. 

\section[Rabi model under BBGS-Darboux transformation]{\label{level2aw}Rabi model under Darboux transformations}

In this Section, we present the theoretical explanation underlying the perturbation effects of atom-field in cavity due to the presence of the classical fields. The perturbation causes the emergence of new potential in which it is perturbed on $y$-axis of Bloch sphere. To do so, it is accomplished through two step of procudures: \textit{first}, the representation of the system is changed from Schr\"{o}dinger equation into one-dimensional Dirac equation. The motivation of this change of representation is because one-dimensional Dirac equation is considered more promising rather than Schr\"{o}dinger equation, since one-dimensional Dirac equation admitting vectors in their potential, instead of Schr\"{o}dinger equation which its potential is in the scalar form only. \textit{Second}, the action of Darboux transform is applied on the Dirac equation, physically meaning that the physical system is perturbed so that the potential and the state of the system can be controllable.
 
We rigorously evaluate BBGS-Darboux transformations in Rabi model, the results substitute into the modified Rabi frequency to obtain the dynamics of Rabi oscillation. The evaluation of BBGS-Darboux transformations in Rabi model involve three Pauli matrices : $\{\sigma_{1},\sigma_{2},\sigma_{3}\}$. The outcomes of this evaluation are : $\it{first}$, the perturbation terms, i.e., the coefficients $\alpha_{i}(t)$ and $\beta_{i}(t)$ which are required to obtain the transformations operator, $B$; $second$, the new Dirac potential of field, $|V_{i}[1](t)|$. Moreover, we also obtain the computational result following the theory; to accomplish it, several methods are needed to simplify the differential equations form and also to obtain the exact solutions of the equations. We find that Homotopy Perturbation Method (HPM) \cite{swilam2009exact} is necessary when solving the coupled-nonlinear Riccatti differential equations problem for  ${\mathcal D (\sigma_{1})}[N=1]$, while for ${\mathcal D(\{\sigma_{2},~\sigma_{3}\})}[N=1]$, the choices produce ordinary differential equations only; however, one needs the Theorem in Ref. \cite{mostowski1964introduction, rabinowitz1993find} to simplify the equations in complex number form by finding their square roots, in case they can be easily simulated.

\subsection[Rabi model in one-dimensional Dirac equation]{\label{level2awa}Rabi model in one-dimensional Dirac equation}

The first procedure, i.e., the change of representation from Schr\"{o}dinger equation into Dirac equation, is given in this Subsection. Rewrite Schr\"{o}dinger equation into one-dimensional stationary Dirac equation of Rabi model
\begin{equation}
\label{eq5}
\hat{h}\Psi =\varepsilon_{0}\Psi.
\end{equation}
%\bibliography{apssamp}% Produces the bibliography via BibTeX.
where $\hat{h}$ = ($i\sigma_{z}\frac{d}{dt}$+$V$), $V=-\hbar \Omega(\sigma^{+}b(t)-\sigma^{-}b^{\dagger}(t))$ and $\varepsilon_{0}$ is $\frac{\hbar \omega_{0}}{2}$. We also use the following assumption $b(t)=b_{0}e^{-i\omega t}$, and \begin{equation}\label{beta1} b_{i}[1](t)=2\beta_{i}(t)-b(t).\end{equation} Because $b_{0}$ is a constant and real, $b_{0}^{\dagger}=b_{0}$, then. 

The Dirac potential in the equation (\ref{eq5})  can be easily changed into $V(t)=i\hbar \Omega b_{0}(\sigma_{x}sin(\omega t)-\sigma_{y}cos(\omega t))$ meaning that the potential is in vector form using Pauli matrices as orthogonal basis. 

Although the idea of representing the physical system in one-dimensional Dirac equation may be better than Schr\"{o}dinger equation, since the potential of Dirac Hamiltonian is in vector form, however it has the drawback: the atomic excitation term in the Hamiltonian is vanished in the new representation. One of justification for this drawback is that the new representation may be valuable to explain the phenomenon in which strong coupling is much greater than atomic excitation as observed in Ref. \cite{Nomura:2010fk}.

\subsection[The action of BBGS-Darboux transformation on Rabi model in one-dimensional Dirac equation]{\label{level2awb}The action of BBGS-Darboux transformation on Rabi model in one-dimensional Dirac equation}

Following the procedure of representation is the action of Darboux transformation on the new representation. Below, the procedure is elucidated deeply.

The one fold BBGS-Darboux transformations change 

\begin{align*}
V(t)&=\sum_{j=1}^{2}\sigma(j)^{j}f_{j}=\sigma(1)^{1}f_{1}(t)+\sigma(2)^{2}f_{2}(t)\\ \nonumber
&=-\hbar \Omega(\sigma^{+}b(t)-\sigma^{-}b^{\dagger}(t))\\ \nonumber
&=i\hbar \Omega b_{0}(\sigma_{x}sin(\omega t)-\sigma_{y}cos(\omega t))\\ \nonumber
\rightarrow \nonumber
V_{i}[t]&=\sum_{j=1}^{2}\sigma(j)^{j}f'_{j}=\sigma(1)^{1}f'_{1}(t)+\sigma(2)^{2}f'_{2}(t)\\ \nonumber
&=-\hbar \Omega(\sigma^{+}b_{i}[1](t)-\sigma^{-}b_{i}^{\dagger}[1](t))\\ \nonumber
&=-i\hbar \Omega (\sigma_{y}(2\beta_{i}(t)-b_{0}cos(\omega t))+\sigma_{x}(b_{0}sin(\omega t))). \nonumber
\end{align*} 

It is clear that the transformations change the initial electromagnetic field over the $y$-direction on the Bloch sphere in the magnitude of $2\beta_{i}(t)$. Due to the transformations, the potential magnitude changes 
\begin{eqnarray}
\label{vpot1}
|V(t)|=(\hbar \Omega b_{0})^{2} \rightarrow ~~~~~~~~~~~~~~~~\nonumber
\\|V_{i}[1](t)|=4\beta_{i}^{2}(t)-4\beta_{i}(t)b_{0}\cos(\omega t)+b_{0}^{2}.
\end{eqnarray}
(One can find by the similar manner that for Jaynes-Cummings model $|V(t)| = (\hbar \Omega)^{2}(n+\frac{1}{2})$. \footnote{The remarkable feature under this representation is that the limit between quantum and classical is simply $|b_{0}|=\sqrt{n+\frac{1}{2}}$})

From intertwining operation $\hat{L}\hat{h}=\hat{h}_{1}\hat{L}$, one can obtain the following equation:
\begin{eqnarray}
\label{meq}
i\sigma_{z}\dot{B}+\hbar \Omega (\sigma^{+}b_{i}[1](t)+\sigma^{-}b_{i}^{\dagger}[1](t)))B-\hbar \Omega (B\sigma^{+}b(t)+ \nonumber
\\B\sigma^{-}b^{\dagger}(t)))-\hbar \Omega (\sigma^{+}\dot{b}(t)+\sigma^{-}\dot{b}^{\dagger}(t)))=0.~~~~~~~~~~~~~
\end{eqnarray}

This is the master equation for the following Subsections. 

\subsubsection[${\mathcal D(\sigma_{1})}$]{\label{sec:level2wp}${\mathcal D(\sigma_{1})}$}

First, we consider ${\mathcal D(\sigma_{1})}$
 to determine $\alpha_{1}(t)$, $\beta_{1}(t)$, and $|V_{1}[1](t)|$. By substituting $B=\alpha_{1}(t)+i(b(t)-\beta_{1}(t))\sigma_{1}$ into the equation (\ref{meq}), it produces the coupled nonlinear Riccati differential equations as following
\begin{subequations}
\label{Riccati1}
\begin{align}
\frac{\dot{\alpha}_{1}(t)}{2\hbar \Omega}-\beta_{1}^2(t)+\beta_{1}(t)(e^{i\omega t}+b_{0}e^{-i\omega t})+b_{0}=0,
\end{align}
\begin{align}
-\dot{\beta}_{1}(t)+i\omega b_{0} e^{-i\omega t}(\hbar \Omega -1)+2\alpha_{1}(t)\hbar \Omega  \nonumber
\\\times(\beta_{1}(t)-cos(\omega t))=0,~~~~~~~~~~~~~~~~~~~~~
\end{align}
\end{subequations}
which are subject to the following initial conditions $\beta_{1}(0)=(\beta_{1})_{0}=b_{0}$ and $\alpha_{1}(0)=(\alpha_{1})_{0}=\frac{i\omega b_{0}(1-\hbar \Omega)}{2 \hbar \Omega(b_{0}-1)}$. In order to obtain the exact solution of these equations, we follow the suggestion for using the Homotopy Perturbation Method (HPM) \cite{swilam2009exact}. 

HPM is considered as a promising tool to solve exactly coupled nonlinear equations since the method successfully reconciles the homotopy theory with perturbation theory \cite{he2003homotopy}. Prior to their amalgamation, it was generally difficult to obtain exact solutions of coupled nonlinear equations using separately homotopy theory or perturbation theory. Most perturbation methods face problems with parameters: the methods admit the existence of a small parameter, in contrary there is no small parameter in frequent nonlinear problems. On the other side, most homotopy methods fail in defining the deformation of most coupled nonlinear differential equations into simpler equations due to their complexities.

As pointed out by Sweilam \textit{et al.}, the \textit{modified} HPM is an effort to cover some drawbacks in conventional HPM. In conventional HPM, the solution is truncated in a series in which often coincides with the Taylor series. The problems arise since the series have very slow convergent rate, therefore the solution under HPM may be inaccurate. For that reason, a new idea in this modified method is proposed by involving Pad\`{e} approximant to enlarge the domain of convergence of the solution in which truncated in the Taylor series.

To do so, the truncated solution in Taylor series as found in conventional HPM is transformed by Laplace transformation, followed by finding the Pad\'{e} approximant, and finished by taking the inverse Laplace transform to obtain a more accurate solution of the problem. We try to apply these similar procedures in order to obtain the exact solutions of the equation (9). The complete algorithms are given in Appendix (A).

Starting from solving the system in the equation (9) by using conventional HPM, one can find that the linear equations in terms of $p$, which is an embedding parameter, of these two equations are
\begin{subequations}
\begin{eqnarray}
p^{0} : \frac{(\dot{\alpha}_{1})_{0}[t]}{2 \Omega  \hbar }=0,~~~~~~~~~~~~~~~~~~~~~~~~~~~~~~~~~~~~~~~~~~~~~
\\-(\dot{\beta}_{1})_{0}[t]=0,~~~~~~~~~~~~~~~~~~~~~~~~~~~~~~~~~~~~~~~~~~~
\end{eqnarray}
\end{subequations}
\begin{subequations}
\begin{eqnarray}
p^{1} :\frac{(\dot{\alpha}_{1})_{1}[t]}{2 \Omega  \hbar }=b_0+(\beta_{1})_{0}[t] ~~~~~~~~~~~~~~~~~~~~~~~~~~~~~~~~~\nonumber
\\ \times(e^{i t \omega }+e^{-i t \omega }b_0)+((\beta_{1})_{0}[t])^2,~~~~~~~~~~~~~~~~~~~~~~~~~~~~~
\end{eqnarray}
\begin{eqnarray}
(\dot{\beta}_1)_{1}[t]=i e^{-i \omega t} \omega  (\Omega  \hbar -1) b_{0}~~~~~~~~~~~~~~~~~~~~~~~~~~~~~~~~\nonumber
\\+2  (\alpha_{1})_{0}[t] \hbar \Omega ((\beta_{1})_{0}[t]- cos[\omega t]),~~~~~~~~~~~~~~~~~~~~~~~
\end{eqnarray}
\end{subequations}
\begin{subequations}
\begin{eqnarray}
p^{2} : \frac{(\dot{\alpha}_{1})_2[t]}{2 \Omega  \hbar }=(e^{i \omega t}+e^{-i \omega t} b_0) (\beta_{1})_{1}[t]~~~~~~~~~~~~~~~~~~\nonumber
\\+2 (\beta_{1})_0[t](\beta_{1})_{1}[t],~~~~~~~~~~~~~~~~~~~~~~~~~~~~~~~~~~~~~~~~
\end{eqnarray}
\begin{eqnarray}
(\dot{\beta}_{1})_{2}[t]=2 \Omega  \hbar  (\alpha_{1}) _{1}[t] ((\beta_{1}) _{0}[t]-cos[\omega t])~~~~~~~~~~~\nonumber
\\ +2 \Omega  \hbar  (\alpha_{1})_{0}[t] (\beta_{1})_1[t].~~~~~~~~~~~~~~~~~~~~~~~~
\end{eqnarray}
\end{subequations}

\begin{figure}[ht!]
\centering
\subfigure[]{
\includegraphics[scale=0.45]{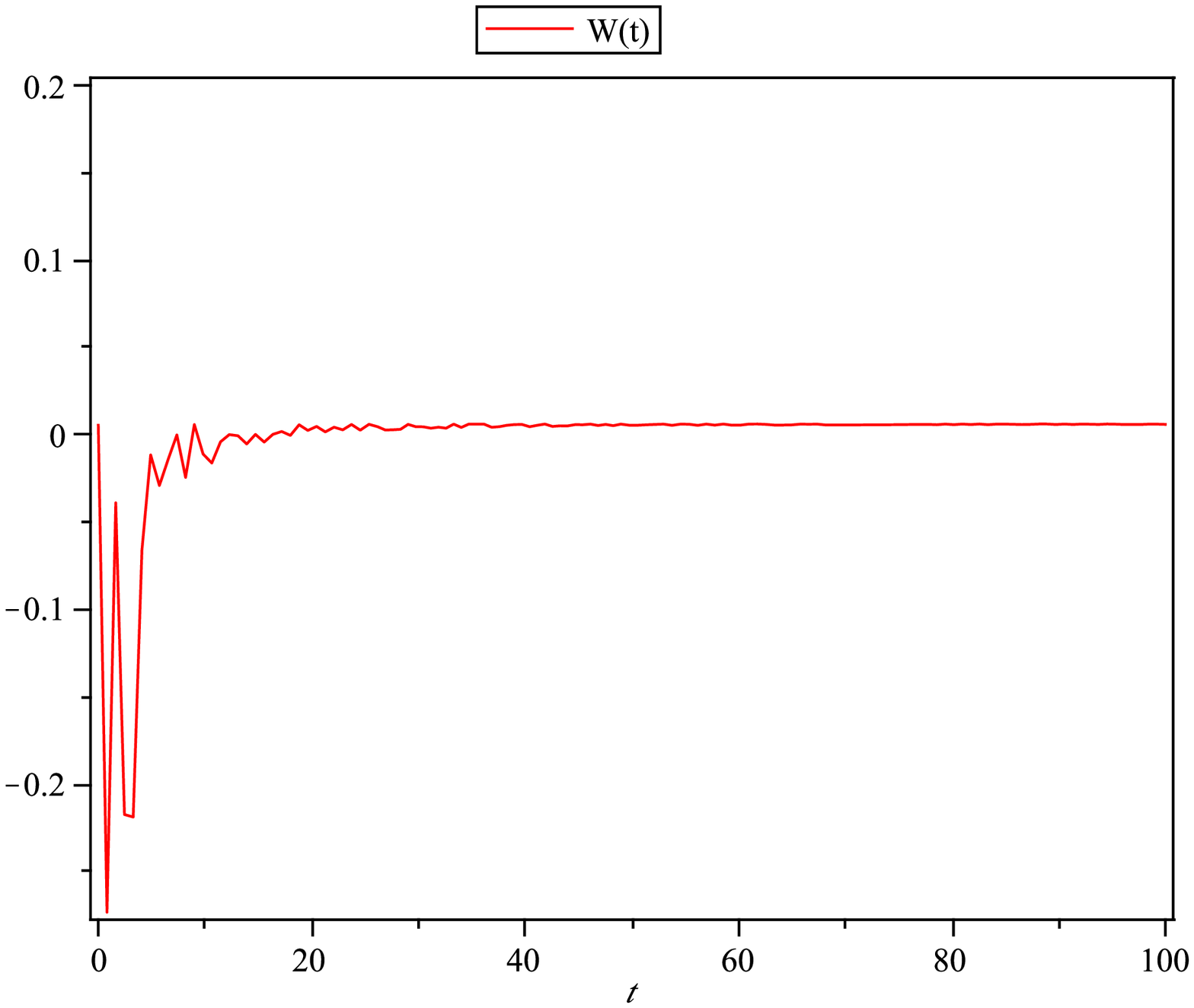}
\label{fig1}
}
\subfigure[]{
\includegraphics[scale=0.45]{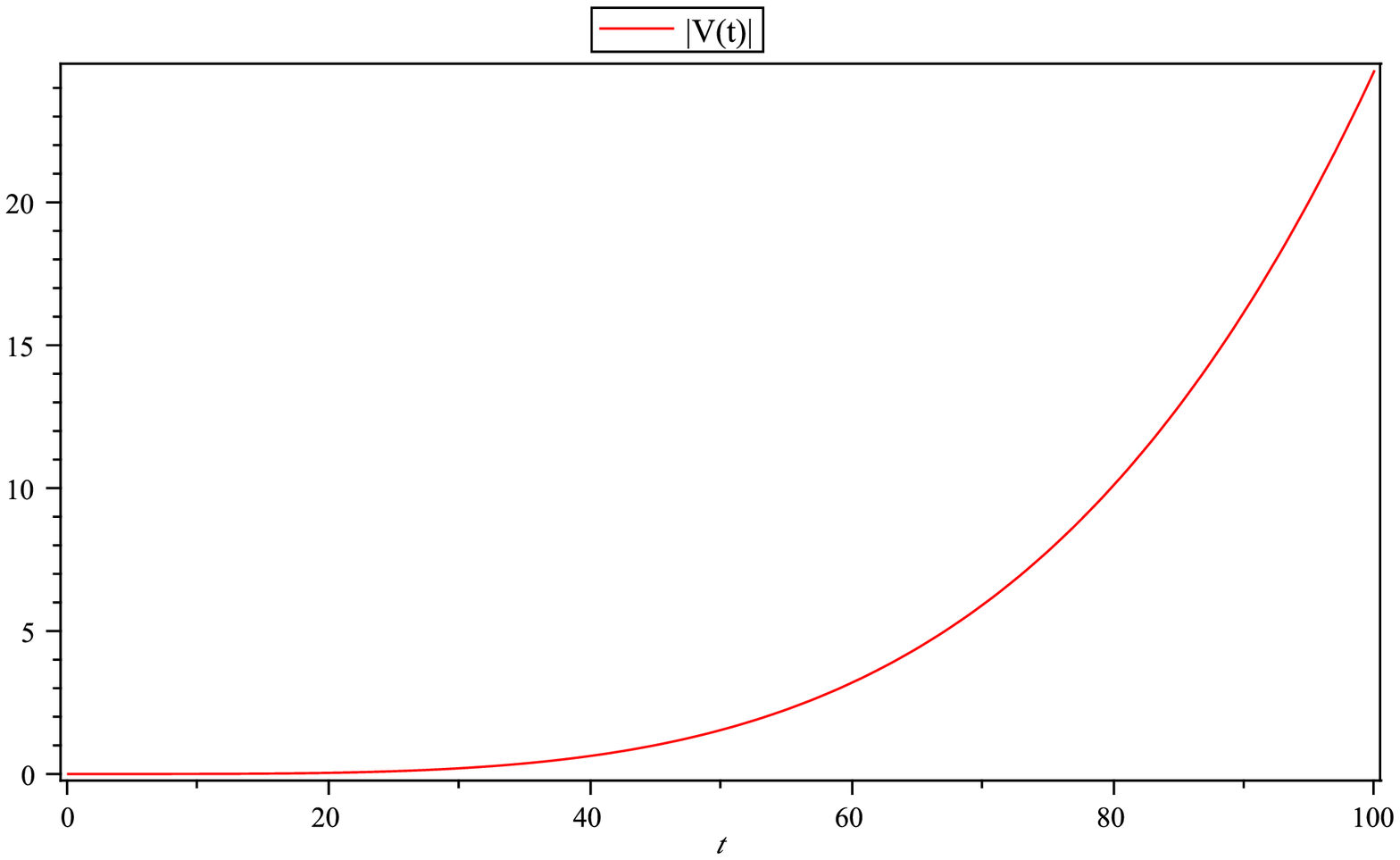}
\label{fig2}
}
\caption[f1]{\label{f1} The time evolution of the population of two states (\ref{fig1}) and the time evolution of Dirac potential (\ref{fig2}) in cavity quantum electrodynamics of ${\mathcal D(\sigma_{1})}$
for $t=100$.}
\end{figure}

Furthermore, the series solution obtained by using the $N$-th order perturbation of HPM can be truncated as following
\\
\begin{widetext}
\begin{subequations}
\label{trunc1}
\begin{eqnarray}
\alpha_{1}(t)\approxeq\sum_{i=0}^{N}(\alpha_{1})_{i}(t)=\frac{i\omega b_{0}(1-\hbar \Omega)}{2 \hbar \Omega(b_{0}-1)}-2\hbar \Omega(b_{0}^{2}-b_{0})(t+i\frac{(1-e^{i\omega t})}{\omega})+{\mathcal O}(t^{2}),
\end{eqnarray}
\begin{eqnarray}
\beta_{1}(t)\approxeq\sum_{i=0}^{N}(\beta_{1})_{i}(t)=b_{0}+(\hbar \Omega-1)((1-e^{-i\omega t})+\frac{i\omega b_{0}}{(b_{0}-1)}(\frac{sin(\omega t)}{\omega}-b_{0}t))+{\mathcal O}(t^{2}).
\end{eqnarray}
\end{subequations}
\end{widetext}
Then, the procedures involving Laplace transforms change $\mathscr{L}[\alpha_{1}(t)]=\hat{\alpha_{1}}(s)$ and $\mathscr{L}[\beta_{1}(t)]=\hat{\beta_{1}}(s)$ in which \textit{s} is replaced by $\frac{1}{t}$. The definition of Laplace transform $\mathscr{L}[\alpha_{1}(t)]=\hat{\alpha_{1}}(s)$ can be found in the Appendix \ref{level0ap}. These procedures yield
\\
\begin{widetext}
\begin{subequations}
\label{Lap1a}
\begin{eqnarray}
\alpha_{1}(t)\approxeq\sum_{i=0}^{N}(\alpha_{1})_{i}(t)=\frac{i\omega b_{0}(1-\hbar \Omega)}{2 \hbar \Omega(b_{0}-1)}t-2\hbar \Omega(b_{0}^{2}-b_{0})t^{2}(1+i\frac{i \omega}{(1-\omega t)})+{\mathcal O}(t^{3}),
\end{eqnarray}
\begin{eqnarray}
\beta_{1}(t)\approxeq\sum_{i=0}^{N}(\beta_{1})_{i}(t)=b_{0}t+(\hbar \Omega-1)t^{2}(\frac{\omega}{1+\omega t}+\frac{i\omega b_{0}}{(b_{0}-1)}(\frac{1}{1+\omega^{2}t^{2}}-b_{0}t^{2}))+{\mathcal O}(t^{3}).
\end{eqnarray}
\end{subequations}
\end{widetext}
Pad\'{e} approximant, $\bigg[\frac{M}{N}\bigg]_{\{\alpha(t),~\beta(t)\}}(t)$ with $\{\textit{M,~N}\}\textgreater$ 0 and \textit{M+N}$\textless~6$, is applied on the equations (\ref{Lap1a}) to enlarge the convergence of these solutions followed by replacing the $\frac{1}{t}$ by $s$. The last step is obtaining the inverse Laplace transforms, $\mathscr{L}[\{\hat{\alpha_{1}}(s),~\hat{\beta_{1}}(s)\}]=\{\alpha_{1}(t),~\beta_{1}(t)\}$, to obtain the true solutions as following
\\
\begin{widetext}
\begin{eqnarray}
\label{v1a}
\alpha_{1}(t)\approx\frac{1}{12} (b_{0}-b_{0}^{2}) \hbar \Omega  (3 \text{i$\omega $} (4 i t \omega +t^2 \omega ^2-2) (\hbar \Omega-1 )+4  \hbar \Omega ^2(b_{0}-1)^2 t (9 i t \omega +t^2 \omega ^2-12))+{\mathcal O}(t^3),
\end{eqnarray}
\begin{eqnarray}
\label{b1}
\beta_{1}(t)\approx\frac{\frac{2 i (2 b_{0}-1) (\hbar \Omega-1)}{\omega }-\frac{i (3 b_{0}-2) e^{-i \omega t} (\hbar \Omega-1)}{\omega }-\frac{i b_{0} e^{i \omega t} (\hbar \Omega-1)}{\omega }-i b_{0}^2 t^2 \omega  (\hbar \Omega-1)+2 (b_{0}-1) t (b_{0}+\hbar \Omega-1)}{2 (b_{0}-1)}+{\mathcal O}(t^{3}).
\end{eqnarray}
\end{widetext}
The Dirac potential of this transformation can be obtained by substituting the equation (\ref{b1}) into the equation (\ref{vpot1}). The plots of Rabi oscillations and potential versus time of this transformation are shown in the figure (\ref{f1}).

For the following Subsections, we find that the differential equations are easier and similar.

\subsubsection[${\mathcal D(\sigma_{3})}$
]{\label{sec:level2yr}${\mathcal D(\sigma_{3})}$
}

Second, we evaluate ${\mathcal D(\sigma_{3})}$
 to obtain $\alpha_{3}(t)$, $\beta_{3}(t)$, and $|V_{3}[1](t)|$. By replacing $B=\alpha_{3}(t)+i(b(t)-\beta_{3}(t))\sigma_{3}$ into the equation (\ref{meq}), it yields
\begin{subequations}
\label{a2a}
\begin{align}
\dot{\beta}_{3}(t)+i\dot{\alpha}_{3}(t)+i\omega b_{0} e^{-i\omega t} = 0,
\end{align}
\begin{align}
2 \hbar \Omega (\alpha_{3}(t)(\beta_{3}(t)-b_{0} cos(\omega t))-i b_{0} \beta_{3}(t)(cos (\omega t)+e^{-i \omega t})~~~~~~~ \nonumber
\\ + i (\beta_{3}(t))^{2} + \frac{i b_{0}}{2}(b_{0}(1+e^{-i2 \omega t})+b_{0} \omega (sin (\omega t)+cos (\omega t))) ~~~~~~~ \nonumber
\\ =0 ~~~~~~~~~~~~~~~~~~~~~~~~~~~~~~~~~~~~~~~~~~~~~
\end{align}
\end{subequations}

\begin{figure}[h!]
\begin{center}
\subfigure[]{
\includegraphics[scale=0.43]{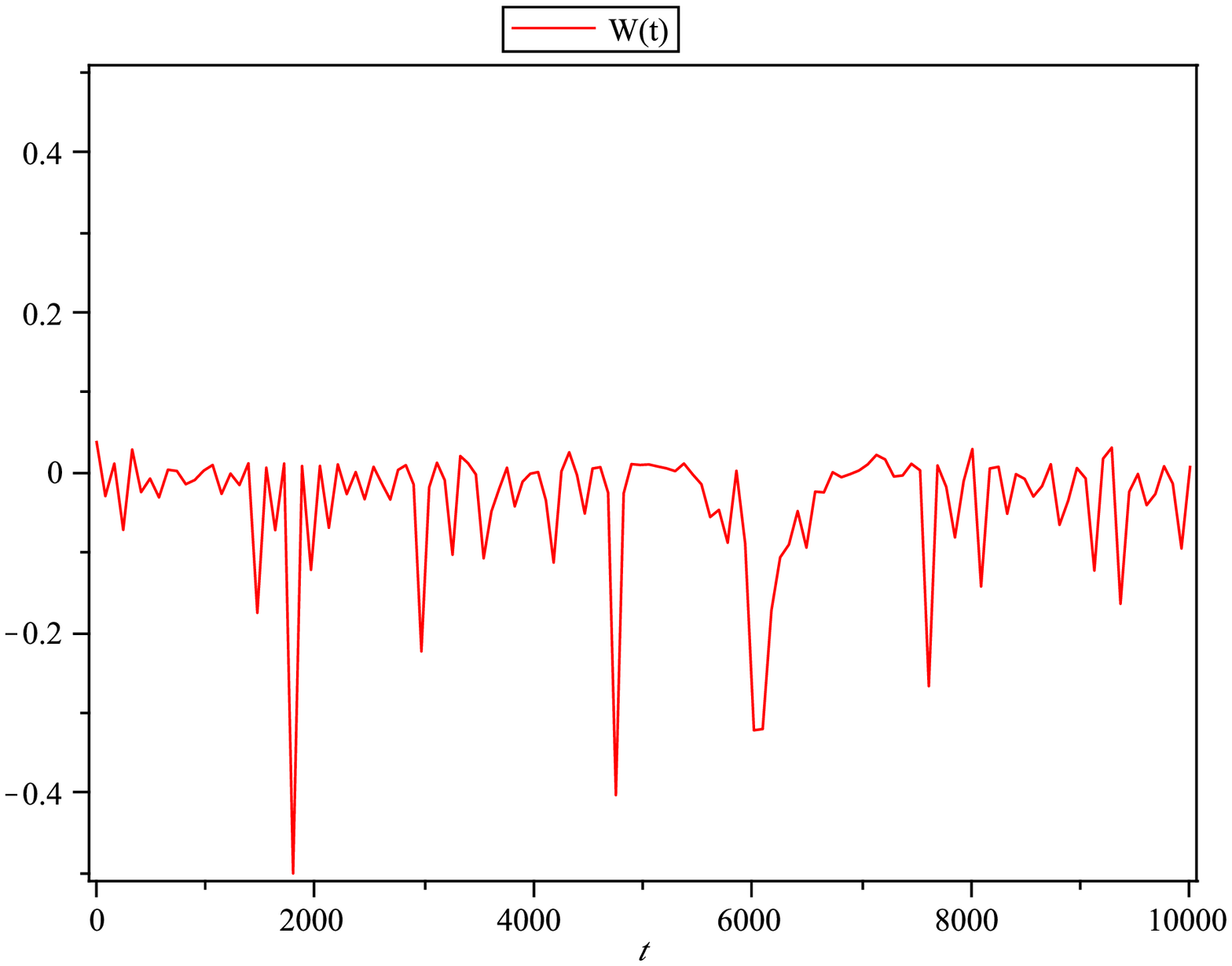}
\label{fg3a}
}
\subfigure[]{
\includegraphics[scale=0.43]{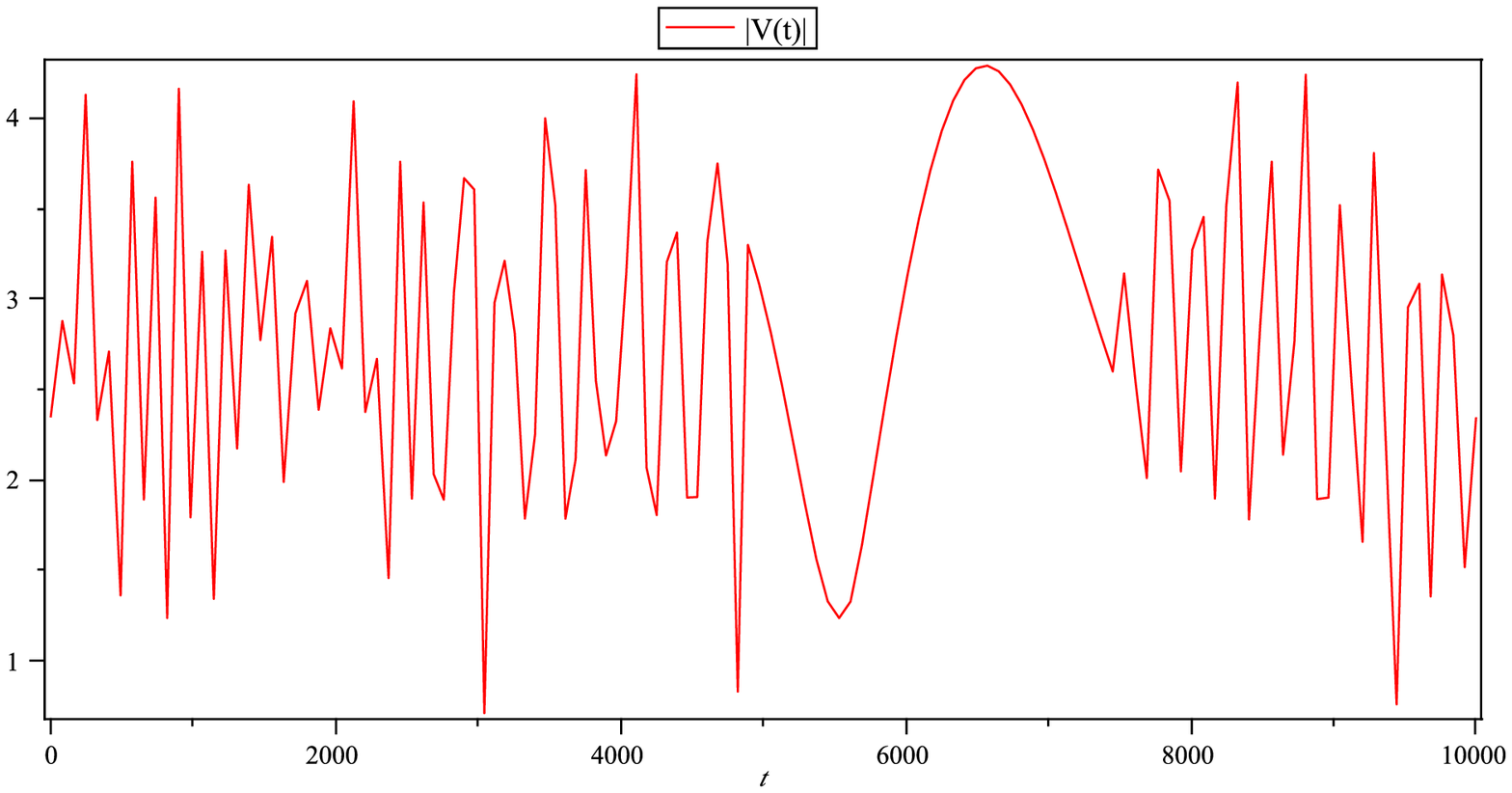}
\label{pot32}
}
\caption[f3]{\label{f3} The time evolution of the population of two states (\ref{fg3a}) and the time evolution of Dirac potential (\ref{pot32}) in cavity quantum electrodynamics of ${\mathcal D(\sigma_{3})}$
for $t=10^{5}$.}
\end{center}
\end{figure}

\subsubsection[${\mathcal D(\sigma_{2})}$
]{\label{sec:level2wr}${\mathcal D(\sigma_{2})}$
}

Third, we examine ${\mathcal D(\sigma_{2})}$
 determine $\alpha_{2}(t)$, $\beta_{2}(t)$, and $|V_{2}[1](t)|$. By changing $B=\alpha_{2}(t)+i(b(t)-\beta_{2}(t))\sigma_{2}$ into the equation (\ref{meq}), it results

\begin{subequations}
\label{a2b}
\begin{align}
2 \hbar \Omega ((\beta_{2}(t))^{2}-\beta_{2}(t)b_{0}e^{-i \omega t}) +i\dot{\alpha}_{2}(t)~~~~~~~~~~~~~~~\nonumber
\\-(i\omega b_{0}e^{-i\omega t}+ \dot{\beta}_{2}(t))=0,~~~~~~~~~~~~~~~~~~~~~
\end{align}
\begin{align}
-\hbar \Omega b_{0} (i\omega)(sin(\omega t)+cos(\omega t)) ~~~~~~~~~~~~~~~\nonumber
\\+ \hbar \Omega \alpha_{2}(t) (2\beta_{2}(t)-b_{0}e^{i\omega t})- \hbar \Omega \alpha_{2}(t) b_{0}e^{-i\omega t}=0.
\end{align}
\end{subequations}

The equations in (\ref{a2a}) and (\ref{a2b}) are ordinary differential equations and they are easy to be solved. However, one needs the theorem as given in Appendix (\ref{level0ac}) to simplify the solutions, in case they are easily modelled.

Thus, from the equations (\ref{a2a}) and (\ref{a2b}), it can be obtained that $\alpha(t)_{(\{2,3\})}$ and $\beta(t)_{(\{2,3\})}$ for ${\mathcal D(\{\sigma_{2},~\sigma_{3}\})}[N=1]$. Since the expression for $\{\alpha(t)_{(\{2,3\})},~\beta(t)_{(\{2,3\})}\}$ and $|V_{(\{2,3\})}[1](t)|$ are too long, it is not necessary to be shown in this paper.

\begin{figure}[h!]
\begin{center}
\subfigure[]{
\includegraphics[scale=0.43]{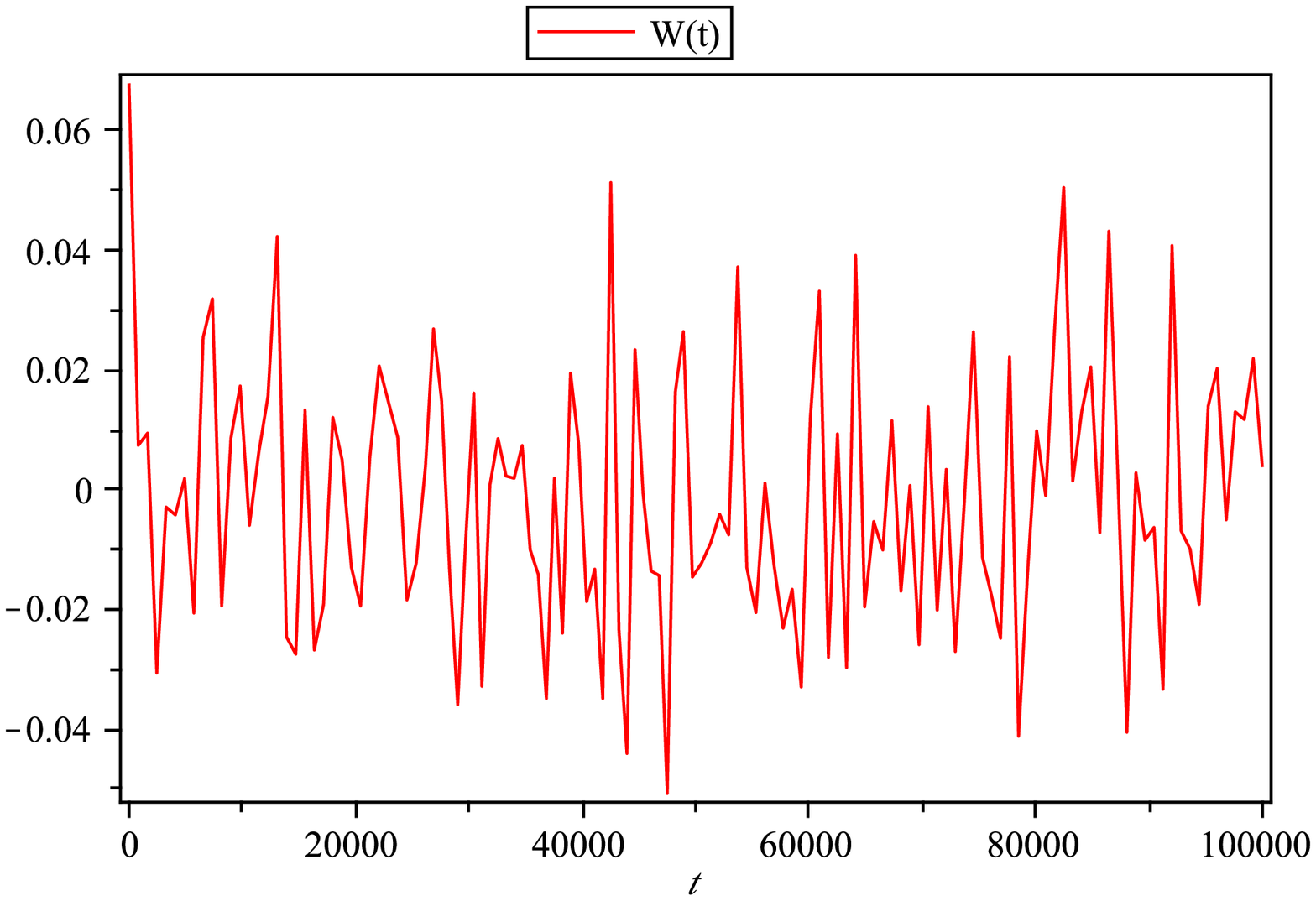}
\label{fig2a}
}
\subfigure[]{
\includegraphics[scale=0.43]{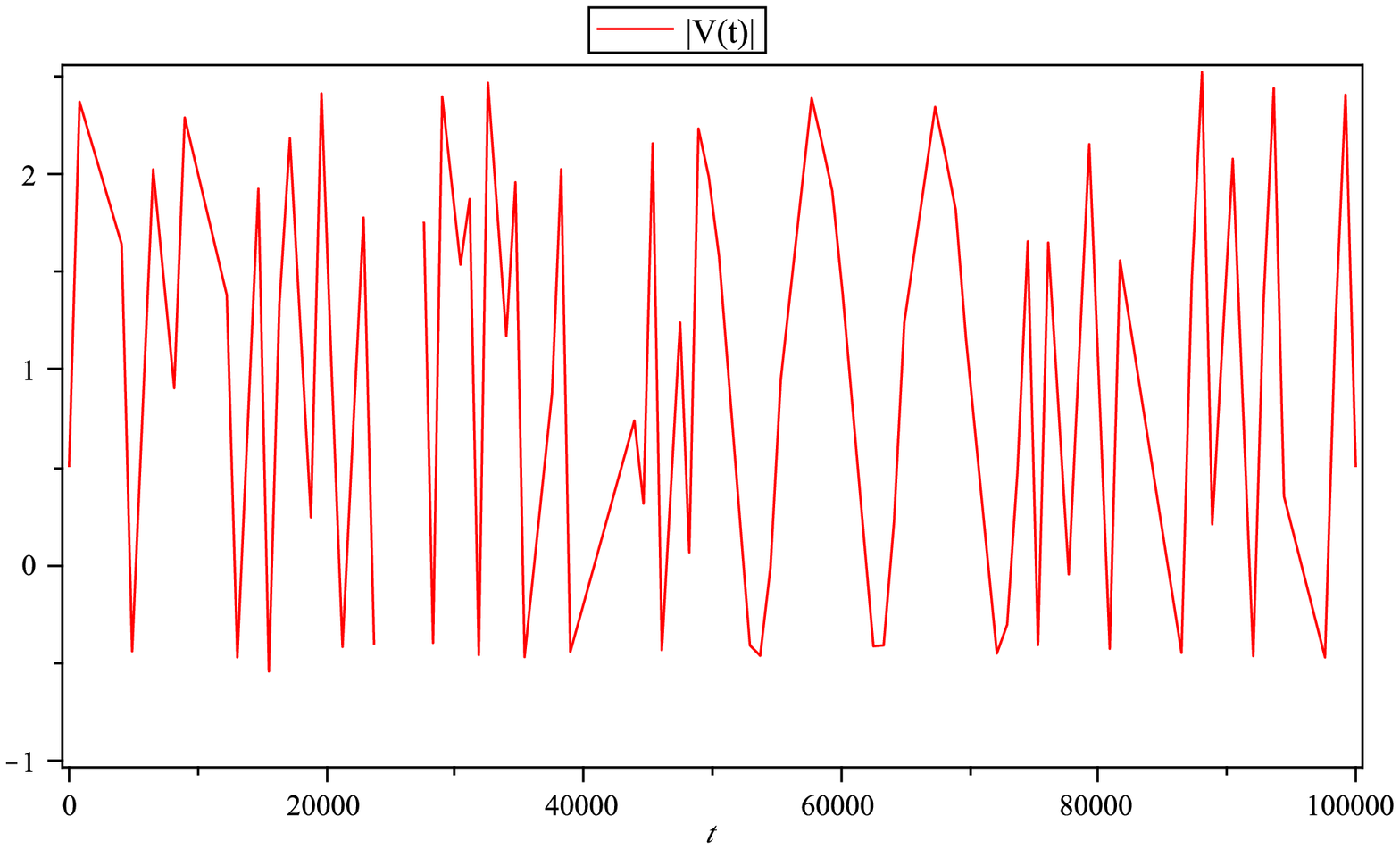}
\label{pot33}
}
\caption[f2]{\label{f2} The time evolution of the population of two states (\ref{fig2a}) and the time evolution of Dirac potential (\ref{pot33}) in cavity quantum electrodynamics of ${\mathcal D(\sigma_{2})}$
for $t=10^{5}$.}
\end{center}
\end{figure}

The potentials can be obtained by substituting those parameters into the equation (\ref{vpot1}). The plots of the Rabi oscillations and the potentials versus time are given in the figures (\ref{f3}) and (\ref{f2}), respectively.

\section[Correlations of the one-dimensional stationary Dirac field potential and the Rabi oscillations]{\label{level6az}Correlations of the one-dimensional stationary Dirac field potential and the Rabi oscillations}

In this Section, we evaluate our results in the previous Section which are believed casting much new light on the underlying physical problem.

The calculations in previous Section result $\beta_{i}(t)$ which are useful to obtain the transformations from $b^{\dagger}b \rightarrow b_{i}^{\dagger}[1]b_{i}[1]$ as defined by the equation (\ref{beta1}). Furthermore, these new classical fields, $\{b_{1}^{\dagger}[1]b_{1}[1], b_{2}^{\dagger}[1]b_{2}[1], b_{3}^{\dagger}[1]b_{3}[1]\}$, are substituted into the \textit{modified} Rabi frequency as expressed by the equation (\ref{mrf}). Consequently, it influences the modified atomic inversion in the equation (\ref{modatin}). The illustration of these schemes can be found on the figure (\ref{conc}) in the Appendix (\ref{level8ao}).

Initially, the atom-field system has the potential, $V(t)$, which the magnitude is $(\hbar \Omega)^{2}|b_{0}|$ in the semiclassical Rabi model and $(\hbar \Omega)^{2}(n+\frac{1}{2})$ for the Jaynes-Cummings model. Then, it is transformed into several types of new potentials.  The results can be divided into two parts : $\it{first}$, the transformation results the parabolic  potential which is performed by ${\mathcal D(\sigma_{1})}$
; $second$, the transformations yield the harmonic oscillator potentials which can be achieved by ${\mathcal D(\{\sigma_{2},~\sigma_{3}\})}$. We provide in the Appendix (\ref{level8ao}) the detailed of computational methods for producing the graphs in this Section.

\subsection[The parabolic potential]{\label{sec:level0bz}The parabolic potential}

The choice of $\sigma_{1}$ corresponds to circumvent the atomic population into the equilibrium condition. 

As can be seen in figure (\ref{fig1}), the absence of Dirac potential causes the atom inclined to occupy the ground state: during the potential is turned off, i.e., for $t~\textless~10$, the first oscillations form wells meaning that initially the atoms occupy ground state, $|g\rangle|n\rangle$, since the wells occupy the negative regime.

Furthermore, the potential is turned on after $t = 10$. It causes that the atomic population in ground state is gradually vanished and followed by coherently sharing the atomic population between the excited- and ground-state due to the presence of the potential. Under the circumstances, the Rabi oscillations are collapsed and extremely damped out to monotonous line. 
 
This phenomenon is similar to the previous results in Ref. \cite{PhysRevA.45.5135} and Ref. \cite{PhysRevA.46.R6801}: due to the classical drive field on resonance, the population is shared coherently between the states $|g,0\rangle$ and $|l,1\rangle$, where $|l,1\rangle=\frac{(|e,0\rangle+|g,1\rangle)}{2}$.

The figure (1(b)) shows that the Rabi oscillations is $\sim \frac{e^{-kt}}{t}$, since the form remembers to Yukawa`s potential \cite{yukawa1935interaction}, thus we name it by $\it{Yukawa-Rabi~oscillations}$. This oscillations can be generated if the Dirac potential is parabolic  as shown in fig (\ref{fig2}). 

The results may be very important for the future of quantum information. The choice of $\sigma_{1}$ causes the coherent share between the excited- and ground-state meaning that in the case that strong interaction of classical field is much greater than atomic excitation in cavity quantum electrodynamics, the parabolic Dirac potential can change the qubit basis from $\sigma_{z}$-eigenbasis into $\sigma_{x}$-eigenbasis.

\subsection[The  harmonic oscillator potentials]{\label{sec:level0bz}The  harmonic oscillator potentials}

The figure (\ref{fg3a}) shows that ${\mathcal D(\sigma_{3})}$ generates the sequences of peaks in negative regime which means the population is driven into the ground state. The sequences of peaks in Rabi oscillations can be created by adjusting fast oscillations of the Dirac potential in the positive regime as shown in the figure (\ref{pot32}). Under this scheme, the sharpness of the peak in atomic inversion is directly proportional to the oscillations speed of Dirac potential: for instance, when $t\sim 2000$, the extreme sharp peak is produced if the fast oscillations of Dirac potential occur. Contrarily, when $t\sim 6000$, the unsharp peak is caused because the oscillations of Dirac potential are smooth and slow.   

The emergence of these peaks in Rabi oscillations and the correlations with the Dirac potentials, may be relevant in the case of \textit{circuit} quantum electrodynamics consisting of two superconducting qubits coupled to an on-chip coplanar waveguide (CPW) and explored by the power dependence of the heterodyne transmission as shown in Ref. \cite{bishop2008nonlinear}. Bishop \textit{et al.} show that the emergences of peaks identified with a multiphoton-transmon qubit transition from the ground state to an excited Jaynes-Cummings state can be accomplished due to the drive power of heterodyne transmission. In contrast with our results, the transition begins to saturate as the drive increases.

Furthermore, the choice of ($\sigma_{2}$) produces Rabi oscillations which are proportional to the oscillations of Dirac potential.

It is shown in the figure (\ref{fig2a}) that initially the atom is in the excited state. The atomic population is gradually driven into the ground state along with the oscillations of potential gradually change from the positive- into the negative-regime. 

This choice demonstrates that it is possible to ensure the coherence of the qubits without the collapse of Rabi oscillations by tuning the Dirac potential in the oscillations form across the positive-negative regimes as shown in figure (\ref{pot33}). This type of Rabi oscillations may be similar to the case of an atom which is initially in the excited state and field initially in a thermal state \cite{knight2005introductory}.
\\
\section[Physical implementation]{\label{sec:level5w}Physical implementation}

The proposed method suggests that the scheme is working under the apparatus in which the electromagnetic fields $\textit{concurrently}$ behave quantumly and classically. The one of possibilities to perform this is by involving the extra device in the cavity in which it generates the field behaving classically. Another possibility, it can be done by involving instrument performing external forces behaving as $\textit{artificial}$ classical field.

In the recent research of photon in cavity, the atom-cavity system can be excited by photon transmission followed by the photon blockade of an optical cavity enclosing one trapped ion in the regime of strong atom-cavity coupling \cite{birnbaum2005photon}. A single atom path is also controllable by the feedback of photon-by-photon \cite{Kubanek:2009qz}. However, it is not clear how photon, the quantum of the electromagnetic field, contributes to control the excitation and decay of the atomic population.

The next constraint is the classical field contributes into the $\textit{modified}$ Rabi oscillations in which the magnitude is $\sim~\sqrt{b^{\dagger}b}$. The possible experiment of classical field driving the atomic population in cavity may relate to the experimental research realizing the Keldysh picture \cite{keldysh1965ionization}. Recent explorations demonstrate that controlling of electronic motion is enabled in ultrafast laser sources in the mid-infrared region \cite{Colosimo:2008ss}. 

\section[Conclusion]{\label{sec:level6}Conclusion}

In this work, we show that the perturbation of atom-field in cavity due to the extra classical electromagnetic field can be used to control the atomic population in cavity. The perturbation theory on the system is described by the application of one fold Darboux transformations for the potential transformations of Rabi model. For simplicity, we use the result of this transformation to obtain the Rabi oscillations transformation by involving classical effect of electromagnetic in the equation. This method shows that it is possible to control  the collapse and resurgence of Rabi oscillations in cavity QED under Darboux transformations. 

The Pauli matrices are the parameters in the BBGS-Darboux transformations to determine the one-dimensional stationary Dirac potential of electromagnetic field which are responsible for controlling the Rabi oscillations. The appropriate choice
 of the parameters may be necessary for elucidating the surprising responses of the oscillations due to the external perturbations as found in Ref.  \cite{bishop2008nonlinear,hennessy2007quantum}.

Based on the results, it is possible to propose an $\textit{open-loop control}$ mechanism for Rabi oscillations under Darboux transformations : the operator ${\mathcal D(\sigma_{i})}$
 is assumed as the controller and the $\sigma_{i}$ is the system input. The initial system is $\{V,\Psi\}$ and the final system is $\{V[N],\Psi[N]\}$. The output variables which are read by the sensor are the new eigenvalues of the new Hamiltonian, $\varepsilon_{N}(\sigma_{i})$. The next challenge is defining an appropriate Darboux transformations so that the theoretical explanation underlying complete $\textit{closed-loop control}$ mechanism, as proposed in Ref. \cite{H.Mabuchi11152002}, can be realized.

For further studies, it may be interesting to investigate the influence of $N$-fold Darboux transformations to the multi-qubits system for various types of quantum system in addition to Bose-Einstein condensation of exciton polaritons\cite{kasprzak2006bose}, Anyons in a weakly interacting system \cite{Weeks:2007uq}, etc. Furthermore, it is also very interesting if the Dirac potential is extended into \textit{n}-dimensions and related to the various quantum systems, since the extension of the potential dimensions can be exploited to describe \textit{quantum} transistor \cite{Trisetyarso:arXiv1003.4590}.

We expect that this paper open the possibility to involve Darboux transformations into the extensive research of quantum information. 

$\textit{Acknowledgments}$ - This work was supported in part by Grant-in-Aid for Scientific Research by MEXT, Specially Promoted Research No. 18001002 and in part by Special Coordination Funds for Promoting Science and Technology. We also would like to thank Prof. Kohei M. Itoh and Rodney Van Meter Ph.D, Toyofumi Ishikawa, Akhtar Waseem, Luis Jou Garcia, and Pierre-Andre Mortemousque for fruitful discussion.

\section[References]{\label{sec:level7}References}

\pagebreak[3]

\appendix
\section[Homotopy Perturbation Methods]{{\label{level0ap}Homotopy Perturbation Methods}}

In this appendix, we provide the algorithm for obtaining exact solution of coupled nonlinear equations  \cite{swilam2009exact}.

\begin{enumerate}
\item Unravel the differential equations using conventional HPM. In this step, the widely known Taylor series is needed to expand the solution of the nonlinear differential equations, $f(x)$, at the point $x=0$. The series reads:
\begin{equation}
\label{taylor1}
f(x)=\sum_{i=0}^{\infty}c_{i}x^{i}= \sum_{i=0}^{\infty}\frac{f^{i}(0)}{i!}x^{i}.
\end{equation}
\item Shortened the series solution by conventional HPM. This step is accomplished by obtaining the embedding parameter of the solutions.
\item Obtain the Laplace transform of the shortened series. The well-known Laplace transform of a function $y(t)$ reads 
\begin{equation}
\mathscr{L}[y(t)]=\hat{y}(s)=\int_{0}^{\infty}y(t)e^{-st} dt.
\end{equation}
\item Acquire the Pad\'{e} approximant of the prior step. The Pad\'{e} approximant is a rational function to approach the Taylor series expansion as best as possible. It gives
\begin{equation}
\bigg[\frac{M}{N}\bigg]_{f(x)}(x)=\frac{\sum\limits_{i=0}^{M}a_{i}x^{i}}{1+\sum\limits_{i=0}^{N}b_{i}x^{i}},
\end{equation}
where $M$ and $N$ are given positive integers.
\item Obtain the inverse Laplace transform. 
\end{enumerate}

\section[Explicit representation for square root of a complex number]{{\label{level0ac}Explicit representation for square root of a complex number}}

Following the Ref. \cite{mostowski1964introduction, rabinowitz1993find}, we provide the theorem to obtain the explicit representation for square root of a complex number.

\textbf{Theorem}. \textit{A complex number 
\begin{equation}
\sqrt{a+ib}
\end{equation}
can be simplified into
\begin{equation}
p+iq
\end{equation}
by defining
\begin{equation}
p=\frac{1}{\sqrt{2}}\sqrt{\sqrt{a^{2}+b^{2}}+a}
\end{equation} 
and
\begin{equation}
q=\frac{sgn}{\sqrt{2}}\sqrt{\sqrt{a^{2}+b^{2}}-a}
\end{equation}
where}

\begin{equation}
\{a,b,p,q \in \Re~\text{and}~(b \neq 0)\}
\end{equation}
\noindent
\textit{and sgn(b)=$\frac{b}{|b|}$ is defined as the sign of b (to be +1 if b $\textgreater$ 0 and -1 if b $\textless$0).}

\section[Computational simulation by ${\mathcal Maple}$]{{\label{level8ao}Computational simulation by \textbf{${\mathcal Maple}$}}}

This Appendix provides the computational methods to generate graphs in the Section (\ref{level6az}). In this work, we use ${\mathcal Maple}.$ The following codes show the codes for modified atomic inversion simulation.

\begin{figure}[h]
\begin{center}
\includegraphics[scale=0.5]{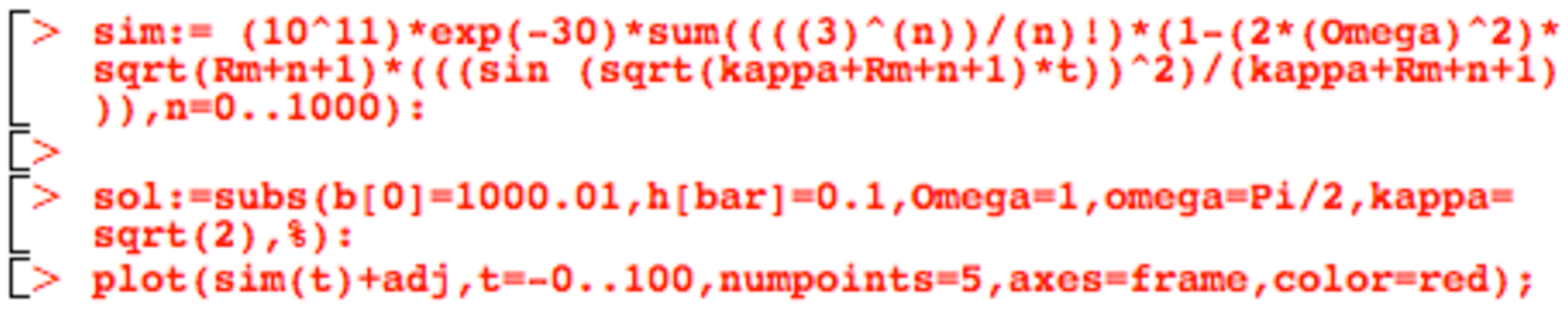}
\caption{\label{maplecode}${\mathcal Maple}$ codes for the \textit{modified} atomic inversion.}
\end{center}
\end{figure}

The meaning of the parameters in the figure (\ref{maplecode}) are \textbf{\textsf{Rm}} is the \textit{modified} Rabi oscillations, \textbf{\textsf{sim}} is the \textit{modified} atomic inversion, \textbf{\textsf{adj}} is the adjustment for atomic inversion in case it is needed, \textbf{\textsf{b[0]}} is the amplitude of classical electromagnetic field, \textbf{\textsf{h[bar]}} is $\hbar$, \textbf{\textsf{kappa}} is $\kappa$, \textbf{\textsf{omega}} is $\omega$, \textbf{\textsf{Omega}} is $\Omega$, and \textbf{\textsf{sol}} is the code for setting the value of the constants.     

Below, the codes of $\{{\mathcal D}(\sigma_{1}),~{\mathcal D}(\sigma_{3}),~{\mathcal D}(\sigma_{2})\}$ solutions are shown, respectively.
\begin{figure}[h]
\begin{center}
\includegraphics[scale=0.5]{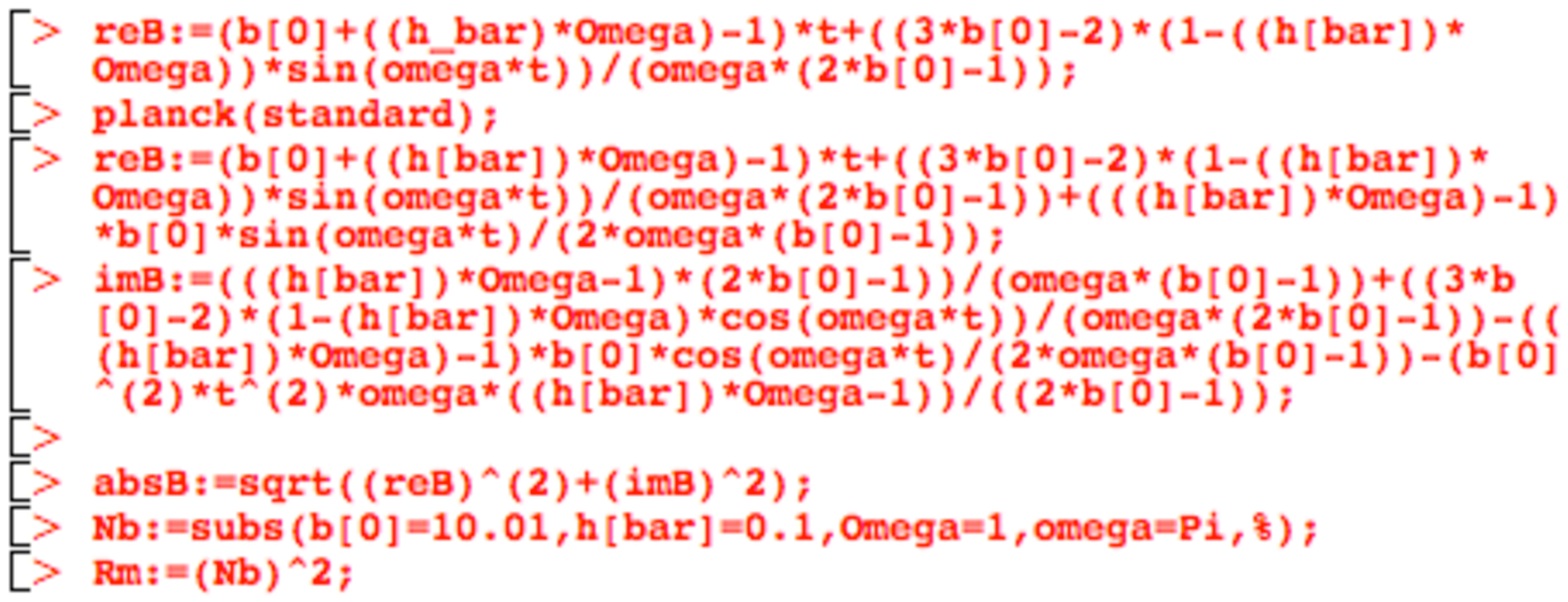}
\caption{\label{sol1}${\mathcal Maple}$ codes for the solution of ${\mathcal D}(\sigma_{1})$.}
\end{center}
\end{figure}

\begin{figure}[h]
\begin{center}
\includegraphics[scale=0.5]{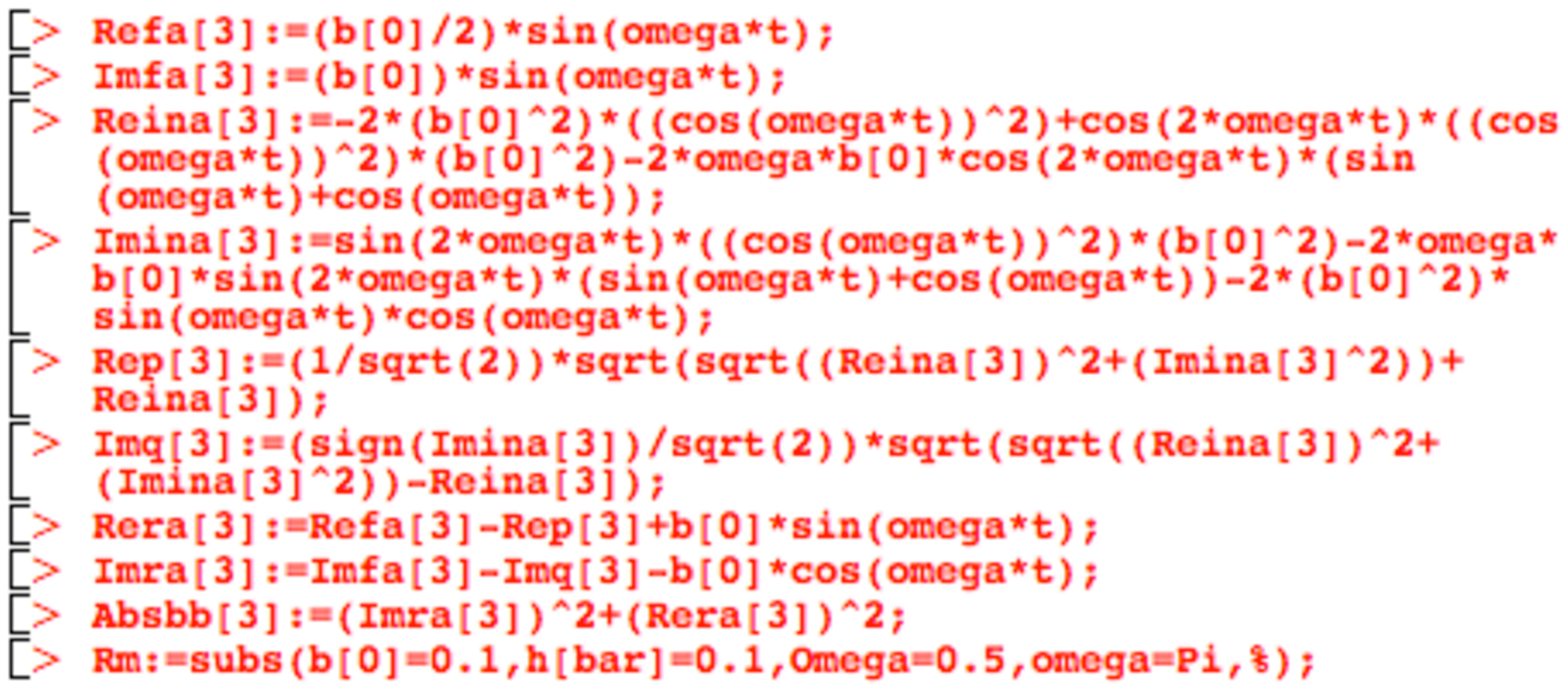}
\caption{\label{sol1}${\mathcal Maple}$ codes for the solution of ${\mathcal D}(\sigma_{3})$.}
\end{center}
\end{figure}

\begin{figure}[h]
\begin{center}
\includegraphics[scale=0.5]{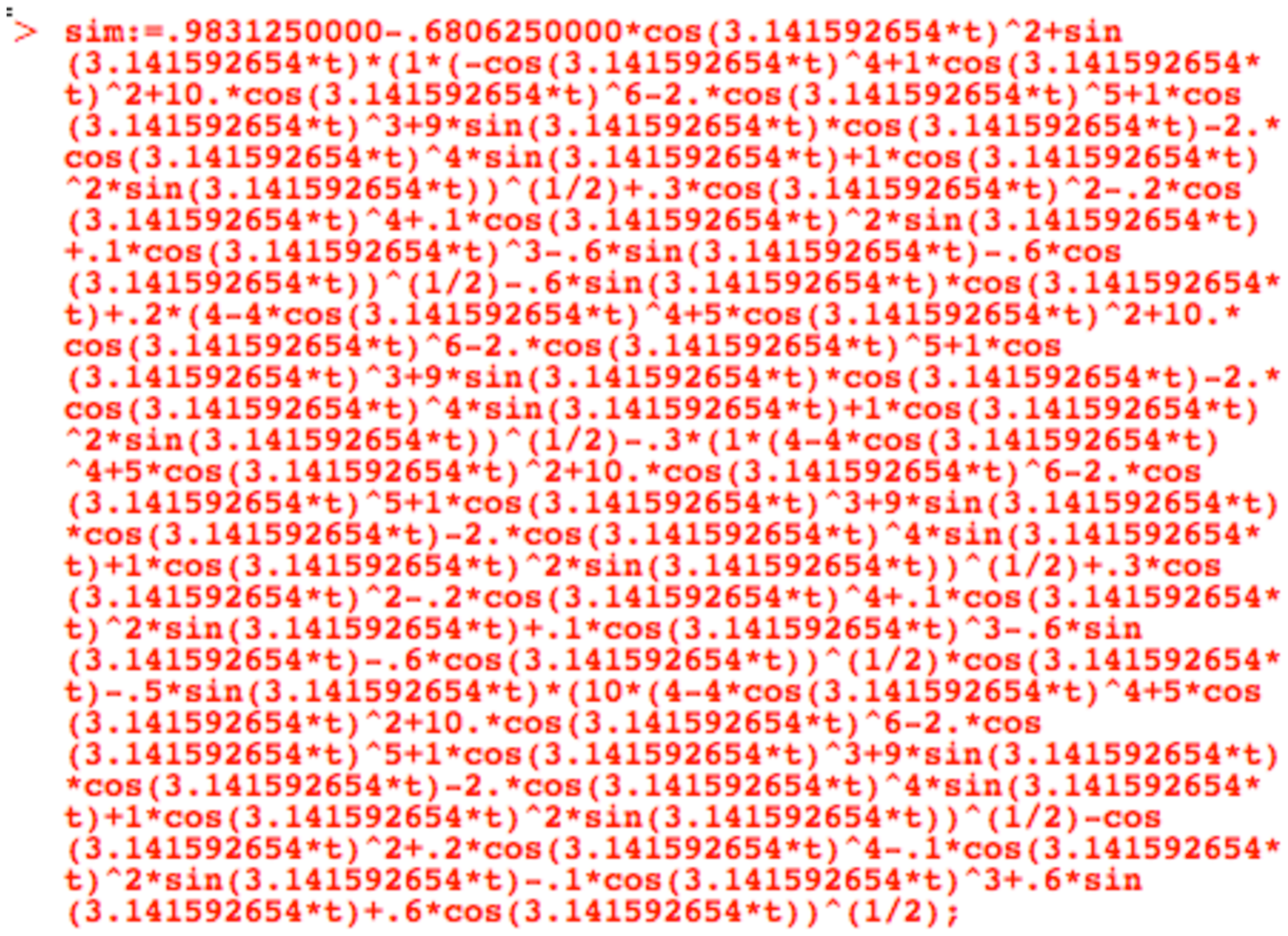}
\caption{\label{sol1}${\mathcal Maple}$ codes for the modified atomic inversion in the case of ${\mathcal D}(\sigma_{2})$.}
\end{center}
\end{figure}

Especially for ${\mathcal D}(\sigma_{2})$, we find the oscillation amplitudes are extremely huge, therefore the amplitude is scaled down in case it is possible to be simulated.

\begin{widetext}
\begin{figure}[h]
\begin{center}
\includegraphics[scale=0.61]{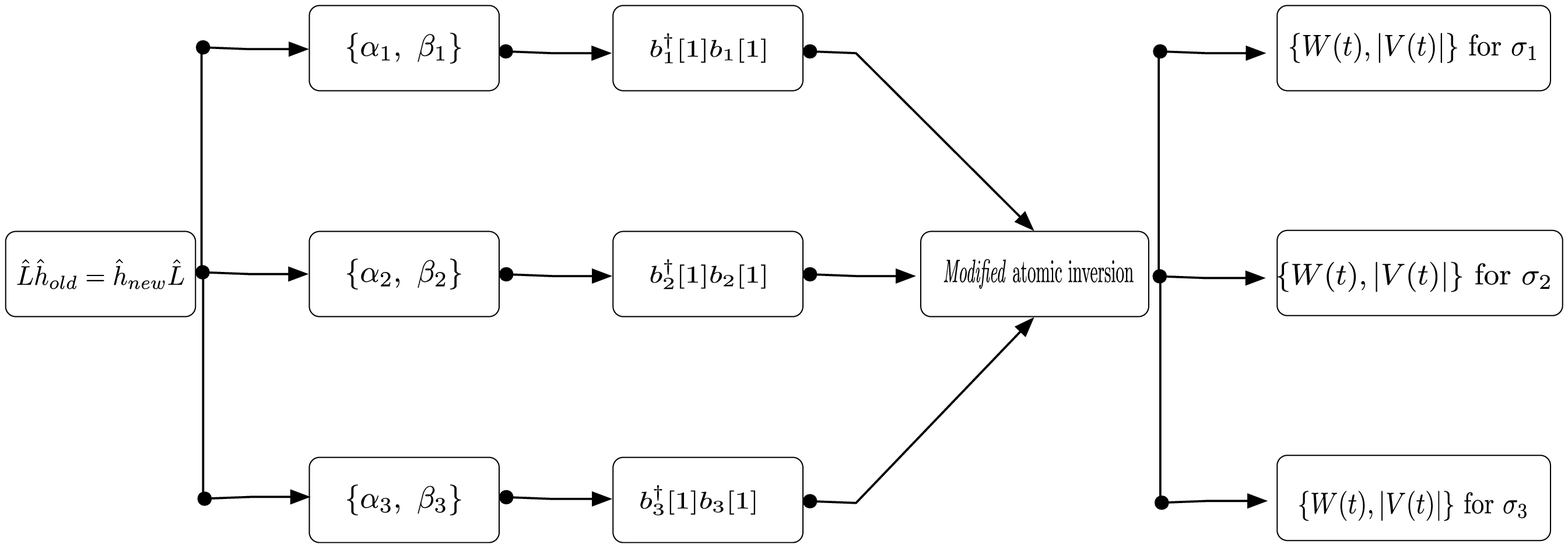}
\caption{\label{conc}The scheme for obtaining the correlations between atomic inversion and Dirac potentials.}
\end{center}
\end{figure}
\end{widetext}

\pagebreak
 
\end{document}